\documentclass[11pt,a4paper]{article} 

\usepackage{jcappub}
\usepackage{amssymb}
\usepackage{amsmath}
\usepackage{hyperref}

\def\Mpl{M_{\rm P}}

\title{A theory of type-II minimally modified gravity}

\author[a]{Antonio De Felice}
\author[a,b]{Andreas Doll}
\author[a,c]{Shinji Mukohyama}

\affiliation[a]{Center for Gravitational Physics, Yukawa Institute for Theoretical Physics,\\ Kyoto University, Kyoto 606-8502, Japan}
\affiliation[b]{Department of Physics and Astronomy,\\
Ruprecht Karls University of Heidelberg, Heidelberg 69117, Germany}
\affiliation[c]{Kavli Institute for the Physics and Mathematics of the Universe (WPI), \\The University of Tokyo Institutes for Advanced Study,\\The University of Tokyo, Kashiwa, Chiba 277-8583, Japan}

\emailAdd{antonio.defelice@yukawa.kyoto-u.ac.jp}
\emailAdd{andreas.doll@yukawa.kyoto-u.ac.jp}
\emailAdd{shinji.mukohyama@yukawa.kyoto-u.ac.jp}

\abstract{We propose a modified gravity theory that propagates only two local gravitational degrees of freedom and that does not have an Einstein frame.  According to the classification in \cite{Aoki:2018brq} (JCAP 01 (2019) 017 [arXiv:1810.01047 [gr-qc]]), this is a type-II minimally modified gravity theory. The theory is characterized by the gravitational constant $G_{\rm N}$ and a function $V(\phi)$ of a non-dynamical auxiliary field $\phi$ that plays the role of dark energy. Once one fixes a homogeneous and isotropic cosmological background, the form of $V(\phi)$ is determined and the theory no longer possesses a free parameter or a free function, besides $G_{\rm N}$. For $V'(\phi) = 0$ the theory reduces to general relativity (GR) with $G_N$ being the Newton's constant and $V=const.$ being the cosmological constant. For $V'(\phi) \ne 0$, it is shown that gravity behaves differently from GR but that GR with $G_{\rm N}$ being the Newton's constant is recovered for weak gravity at distance and time scales sufficiently shorter than the scale associated with $V(\phi)$. Therefore this theory provides the simplest framework of cosmology in which deviations from GR can be tested by observational data. }


\arxivnumber{2004.12549}

\notoc

\begin{document}
\begin{flushright} {\footnotesize YITP-20-55, IPMU20-0040}  \end{flushright}
\vspace{-0.8cm}
\maketitle
\flushbottom

\section{Introduction} \label{sec:intro}
Gravity remains the least understood among the four forces in nature. For example, we know neither the physical origin of the accelerated expansion of the universe at late time nor the physical mechanism that stabilizes the enormous hierarchy between the scale of the accelerated expansion and the Planck scale. These two faces of the so called cosmological constant problem motivate us to speculate various possibilities for modification of gravity at long distance and time scales. Yet, given the fact that general relativity (GR) successfully explains many experimental and observational data such as those from solar system experiments and gravitational waves observations, we do not expect nature to allow large modification except at cosmological scales. Since extra propagating degrees of freedom tend to introduce modifications at various scales, it is interesting to consider minimally modified gravity (MMG) theories~\cite{Lin:2017oow}, i.e. modified gravity theories with two local gravitational degrees of freedom, as possible candidate theories of the origin of the late time accelerated expansion of the universe.

In \cite{Aoki:2018brq} all MMG theories were classified into type-I and type-II. A type-I MMG theory is a theory with two local gravitational degrees of freedom that can be recasted as GR with a non-trivial coupling to matter by a change of variables. In other words, a type-I MMG has an Einstein frame. Here, an Einstein frame is a frame in which all GR solutions are solutions of the theory under consideration at least locally if the matter is minimally coupled to the metric. Therefore, in type-I MMG theories, gravity is modified due to non-trivial matter coupling. On the other hand, a type-II MMG theory is a theory with two local gravitational degrees of freedom that does not have an Einstein frame.

As pointed out in \cite{Aoki:2018zcv} and then elaborated in \cite{Aoki:2018brq}, all type-I MMG theories can be constructed by performing general canonical transformations to GR in the Arnowitt-Deser-Misner (ADM) formalism and then gauge-fixing (part of) the diffeomorphism before adding matter fields. While \cite{Lin:2017oow,Carballo-Rubio:2018czn,Mukohyama:2019unx} found several explicit examples of MMG, most (if not all) of them are of type-I \cite{Carballo-Rubio:2018czn}. Indeed, in the absence of matter, constraint equations in most (if not all) of them are equivalent to those in GR and thus their Hamiltonians are equivalent to that of GR up to redefinition of Lagrange multipliers. One can also consider canonical transformations that involve not only the metric components and their conjugate momenta but also the stress energy tensor of matter fields to construct yet another kind of type-I MMG theories. A simple example of this kind was considered in \cite{Feng:2019dwu}. The cuscuton~\cite{Afshordi:2006ad} is also a theory of MMG (see subsection 3.4 of \cite{Lin:2017oow}, \cite{Iyonaga:2018vnu}, subsection 3.B of \cite{Mukohyama:2019unx} and section IV of \cite{Gao:2019twq} for generalization). In order to judge whether it is of type-I or type-II, one needs to perform a general canonical transformation in the unitary gauge description to see if it allows for an Einstein frame or not. As far as the authors know, this is an open question. As for type-II MMG, an obvious example is the minimal theory of massive gravity \cite{DeFelice:2015hla,DeFelice:2015moy,Bolis:2018vzs,DeFelice:2018vza}, in which the tensor graviton has a non-vanishing mass and thus its dispersion relation is different from that in GR with or without matter.

The purpose of the present paper is to find yet another type-II MMG theory that is theoretically consistent and observationally viable. For this purpose, in section \ref{sec:Hamiltonian} we start with GR in the ADM formalism and perform a canonical transformation that mixes the spatial metric, the lapse function and their conjugate momenta. We then gauge fix the time diffeomorphism and add a cosmological constant in the frame after the transformation. Instead of adding matter in this frame, we perform the inverse canonical transformation to go back to the original frame, which is no longer an Einstein frame because of the additional term induced by the cosmological constant in the other frame. We can then add matter fields to the frame after the inverse canonical transformation. After constructing the Hamiltonian formulation of the theory in this way, we perform a Legendre transformation to obtain the Lagrangian formulation of the theory in section \ref{sec:Lagrangian}. In section \ref{sec:degrees} we then confirm that the number of local gravitational degrees of freedom is two at the fully nonlinear level. In section \ref{sec:cosmology} we study a homogeneous and isotropic cosmological background and linear perturbations around it. In particular, we show that the only free function in the theory can be reconstructed from the cosmological background, leaving the gravitational constant $G_{\rm N}$ as the only parameter. We also show that gravity behaves differently from general relativity (GR) in general but that GR with $G_{\rm N}$ being the Newton's constant is recovered for weak gravity at distance and time scales sufficiently shorter than the scale associated with the late time cosmology. Finally, section \ref{sec:summary} is devoted to a summary of the paper and some discussions.

\section{Hamiltonian formulation}
\label{sec:Hamiltonian}
We propose here a type-II theory of MMG. The idea is simple, and it is based on the following points.
\begin{enumerate}
\item We start from the Hamiltonian of General Relativity written in the ADM formalism.
\item We perform a canonical transformation to a new frame via a generating functional which depends on new variables and old momenta.
\item We introduce a cosmological constant in the new frame.
\item We add a gauge fixing term, in order to keep the theory minimal, i.e.\ with only two local physical degrees of freedom in the gravity sector.
\item We perform the inverse canonical transformation to go back to the original frame, but the theory has now changed.
\item We add standard matter fields in the original frame but inside a theory which is not GR any longer.
  \end{enumerate}

We now follow step by step the above given algorithm. Therefore, we start by considering the Hamiltonian of General Relativity written in the ADM variables and then we perform a canonical transformation. Namely, we begin with the $4$-dimensional metric of the form
\begin{equation}
 g_{\mu\nu}dx^{\mu}dx^{\nu} = -N^2dt^2 + \gamma_{ij}(dx^i+N^idt)(dx^j+N^jdt)\,,
  \label{eqn:metric-ADM}
\end{equation}
and 
\begin{equation}
H_{{\rm tot}}=\int d^{3}x[N\mathcal{H}_{0}(\gamma,\pi)+N^{i}\mathcal{H}_{i}(\gamma,\pi)+\lambda\pi_{N}+\lambda^{i}\pi_{i}]\,,
\end{equation}
where $\lambda$ and $\lambda^i$ are Lagrange multipliers, and 
\begin{eqnarray}
\mathcal{H}_{0} & = & \frac{2}{\Mpl^{2}\sqrt{\gamma}}\left(\gamma_{ik}\gamma_{jl}-\frac{1}{2}\,\gamma_{ij}\gamma_{kl}\right)\pi^{ij}\pi^{kl}-\frac{\Mpl^{2}\sqrt{\gamma}}{2}\,R(\gamma)\,, \label{eqn:H0-GR} \\ 
\mathcal{H}_{i} & = & -2\sqrt{\gamma}\gamma_{ij}D_{k}\!\left(\frac{\pi^{jk}}{\sqrt{\gamma}}\right)\,, \label{eqn:Hi-GR}
\end{eqnarray}
are the Hamiltonian constraint and the momentum constraint, respectively. Here, $D_k$ is the spatial covariant derivative compatible with the metric $\gamma_{ij}$. 
Then we introduce the generating functional as 
\begin{equation}
F=F(\mathfrak{N},\mathfrak{N}^{i},\Gamma_{ij},\pi_{N},\pi_{i},\pi^{ij})\,,
\end{equation}
which can be written schematically as $F=F({\rm new\ }\tilde q,{\rm old\ } p)$, so that $q=-\frac{\partial F}{\partial p}$, and $\tilde p=-\frac{\partial F}{\partial\tilde q}$. 
In this paper, for the sole purpose of simplicity, we restrict our consideration to the following form of $F$:
\begin{equation}
F=-\int d^{3}x\,[\Mpl^{2}\sqrt{\Gamma}f(\phi,\psi)+\mathfrak{N}^{i}\,\pi_{i}]\,,
\end{equation}
where we have introduced the quantities
\begin{equation}
\phi=\frac{1}{\Mpl^{2}\sqrt{\Gamma}}\,\pi^{ij}\,\Gamma_{ij}\,,\qquad\psi=\frac{1}{\Mpl^{2}\sqrt{\Gamma}}\,\pi_{N}\,\mathfrak{N}\,.
\end{equation}
At this level $\phi$ and $\psi$ are just shortcuts for the expressions written above. Later on, we will promote them to be three dimensional auxiliary scalar fields.

Then for this chosen generating functional, we find the following transformations:
\begin{eqnarray}
\Pi_{\mathfrak{N}} & = & -\frac{\delta F}{\delta\mathfrak{N}}=\Mpl^{2}\sqrt{\Gamma}f_{\psi}\,\frac{1}{\Mpl^{2}\sqrt{\Gamma}}\,\pi_{N}=f_{\psi}\,\pi_{N}\,,\\
\Pi_{i} & = & -\frac{\delta F}{\delta\mathfrak{N}^{i}}=\pi_{i}\,,\\
\Pi^{ij} & = & -\frac{\delta F}{\delta\Gamma_{ij}}=\frac{1}{2}\Mpl^{2}\,\sqrt{\Gamma}\,\Gamma^{ij}\,f+\Mpl^{2}\sqrt{\Gamma}f_{\phi}\left(\frac{1}{\Mpl^{2}\sqrt{\Gamma}}\,\pi^{ij}-\frac{1}{\Mpl^{2}(\sqrt{\Gamma})^{2}}\,\pi^{lk}\,\Gamma_{lk}\,\frac{1}{2}\sqrt{\Gamma}\,\Gamma^{ij}\right)\nonumber \\
 &  & {}+\Mpl^{2}\sqrt{\Gamma}f_{\psi}\left(-\frac{1}{\Mpl^{2}(\sqrt{\Gamma})^{2}}\,\pi_{N}\,\mathfrak{N}\,\frac{1}{2}\sqrt{\Gamma}\,\Gamma^{ij}\right)\nonumber \\
 & = & f_{\phi}\,\pi^{ij}+\frac{\Mpl^{2}}{2}\,\sqrt{\Gamma}\,\Gamma^{ij}\,(f-f_{\phi}\phi-f_{\psi}\psi)\,,\\
\gamma_{ij} & = & -\frac{\delta F}{\delta\pi^{ij}}=\Mpl^{2}\sqrt{\Gamma}f_{\phi}\,\frac{1}{\Mpl^{2}\sqrt{\Gamma}}\,\Gamma_{ij}=f_{\phi}\,\Gamma_{ij}\,,\\
N & = & -\frac{\delta F}{\delta\pi_{N}}=f_{\psi}\,\mathfrak{N}\,,\\
N^{i} & = & -\frac{\delta F}{\delta\pi_{i}}=\mathfrak{N}^{i}\,,
\end{eqnarray}
where we have called $f_{\phi}=\partial f/\partial\phi$, and $f_{\psi}=\partial f/\partial\psi$.
Therefore we can also write
\begin{eqnarray}
\psi & = & \frac{1}{\Mpl^{2}\sqrt{\Gamma}}\,\pi_{N}\,\mathfrak{N}=\frac{f_{\phi}^{3/2}}{\Mpl^{2}\sqrt{\gamma}}\,\frac{\pi_{N}}{f_{\psi}}\,N\,,\\
\phi & = & \frac{1}{\Mpl^{2}\sqrt{\Gamma}}\,\pi^{ij}\,\Gamma_{ij}=\frac{f_{\phi}^{3/2}}{\Mpl^{2}f_{\phi}\sqrt{\gamma}}\,\pi^{ij}\,\gamma_{ij}=\frac{f_{\phi}^{1/2}}{\Mpl^{2}\sqrt{\gamma}}\,\pi^{ij}\,\gamma_{ij}\,.
\end{eqnarray}

After performing the canonical transformation at the level of the Hamiltonian, and on promoting $\phi$ and $\psi$ to be two independent three dimensional scalar fields, we have
\begin{eqnarray}
H_{{\rm tot}} & = & \int d^{3}x[\mathfrak{N}f_{\psi}\mathcal{H}_{0}(\Gamma,\Pi,\phi,\psi)+\mathfrak{N}^{i}\mathcal{H}_{i}(\Gamma,\Pi,\phi,\psi)+\tilde{\lambda}\Pi_{\mathfrak{N}}+\lambda^{i}\Pi_{i}+f_\phi^{3/2}\sqrt{\Gamma}\lambda_{C}\,C(\Gamma,\Pi,\phi,\psi)\nonumber \\
 &  & {}+f_\phi^{3/2}\sqrt{\Gamma}\lambda_{D}\mathcal{D}(\Gamma,\Pi,\phi,\psi)+\lambda_{\phi}\pi_{\phi}+\lambda_{\psi}\pi_{\psi}+f_\phi^{3/2}\sqrt{\Gamma}\lambda_{{\rm gf}}^{i}\partial_{i}\phi+\Mpl^{2}\mathfrak{N}\sqrt{\Gamma}\tilde{\Lambda}]\,,\label{eq:Hamil_one}
\end{eqnarray}
where  
\begin{eqnarray}
  C&=&\phi- \frac{1}{\Mpl^{2}\sqrt{\Gamma}}\,\Gamma_{ij}\,\frac1{f_\phi}\left[\Pi^{ij}-\frac{\Mpl^{2}}{2}\,\sqrt{\Gamma}\,\Gamma^{ij}\,(f-f_{\phi}\phi-f_{\psi}\psi)\right]\,,\\
  \mathcal{D}&=& \psi- \frac{1}{\Mpl^{2}\sqrt{\Gamma}}\,\frac{\Pi_{\mathfrak N}}{f_\psi}\,\mathfrak{N}\,.
\end{eqnarray}
Notice that in eq.\ (\ref{eq:Hamil_one}), we have added a gauge-fixing term and a cosmological constant term. The presence of both are necessary to: 1) obtain a theory different from GR, and: 2) to keep the same degrees of freedom as in GR.

On using the inverse transformation we find
\begin{eqnarray}
H_{{\rm tot}} = \int d^{3}x & & \left[N\mathcal{H}_{0}(\gamma,\pi)+N^{i}\mathcal{H}_{i}(\gamma,\pi)+\lambda\,\pi_{N}+\lambda^{i}\pi_{i}+\sqrt{\gamma}\lambda_{C}\left(\phi-\frac{f_{\phi}^{1/2}}{\Mpl^{2}}\,\frac{\pi^{ij}}{\sqrt{\gamma}}\,\gamma_{ij}\right)\right.\nonumber \\
 &  & {}+\left.\sqrt{\gamma}\lambda_{D}\left(\psi-\frac{f_{\phi}^{3/2}}{\Mpl^{2}f_{\psi}}\,\frac{\pi_{N}}{\sqrt{\gamma}}\,N\right)+\lambda_{\phi}\pi_{\phi}+\lambda_{\psi}\pi_{\psi} \right. \nonumber \\
 &  & {}+\left.\sqrt{\gamma}\,\lambda_{{\rm gf}}^{i}\,\partial_{i}\phi+\frac{\Mpl^{2}}{f_{\psi}f_{\phi}^{3/2}}\,N\sqrt{\gamma}\tilde{\Lambda}\right]\,. \label{eqn:Htot_pre}
\end{eqnarray}
Therefore the type-II MMG theory discussed in this paper is exactly defined by the last two terms of the Hamiltonian (\ref{eqn:Htot_pre}). In fact, without these, the theory would be equivalent to GR.

The following two primary constraints 
\begin{eqnarray}
\pi_{N} & \approx & 0\,,\\
\psi & \approx & \frac{f_{\phi}^{3/2}}{\Mpl^{2}f_{\psi}}\,\frac{\pi_{N}}{\sqrt{\gamma}}\,N\,,
\end{eqnarray}
automatically lead to the constraint
\begin{equation}
\psi\approx0\,.
\end{equation}
We also have another primary constraint
\begin{equation}
\pi_{\psi}\approx0\,,
\end{equation}
so that we can eliminate the pair $(\psi,\pi_{\psi})$ from the dynamical
variables. In this case we can expand $f$ as a function of $\phi$
and $\psi$ with respect to $\psi$, as in 
\begin{equation}
f(\phi,\psi)=F(\phi)+f_{1}(\phi)\,\psi+\mathcal{O}(\psi^{2})\,,
\end{equation}
and the constraint $\psi\approx 0$ allows one to stop the expansion at the linear order in $\psi$, as any higher order terms would not contribute to the Hamiltonian or Poisson brackets among constraints. Therefore we find
\begin{eqnarray}
H_{{\rm tot}} & = & \int d^{3}x\!\left[N\mathcal{H}_{0}(\gamma,\pi)+N^{i}\mathcal{H}_{i}(\gamma,\pi)+\lambda\,\pi_{N}+\lambda^{i}\pi_{i}+\sqrt{\gamma}\lambda_{C}\!\left(\phi-\frac{f_{0}^{1/2}}{\Mpl^{2}}\,\frac{\pi^{ij}}{\sqrt{\gamma}}\,\gamma_{ij}\right)\right.\nonumber \\
 &  & \left.{}+\lambda_{\phi}\pi_{\phi}+\sqrt{\gamma}\lambda_{{\rm gf}}^{i}\,\partial_{i}\phi+\frac{1}{f_{1}f_{0}^{3/2}}\,N\sqrt{\gamma}\Mpl^{2}\tilde{\Lambda}\right],
\end{eqnarray}
where $\mathcal{H}_{0}$ and $\mathcal{H}_{i}$ are the standard GR expressions (\ref{eqn:H0-GR})-(\ref{eqn:Hi-GR}), and we have defined 
\begin{equation}
f_{0}\equiv\frac{dF}{d\phi}\,.
\end{equation}

At this point, supposing that we want to study the branch of solution for which GR limit is equivalent to $C\to0$, we can make a field redefinition
\begin{eqnarray}
\phi & = & f_{0}^{1/2}\,\bar{\phi}\,,\\
\frac{f_{0}^{1/2}}{\Mpl^{2}}\,\lambda_{C} & = & \bar{\lambda}_{C}\,,\\
\frac{\Mpl^{2}\tilde{\Lambda}}{f_{1}f_{0}^{3/2}} & = & \Mpl^{2}V(\bar{\phi})\,.
\end{eqnarray}
Then, on redefining $\lambda_{\phi}$ and $\lambda_{{\rm gf}},$ and
removing everywhere the bar for simplicity, we obtain
\begin{eqnarray}
H_{{\rm tot}} & = & \int d^{3}x\!\bigg[N\mathcal{H}_{0}(\gamma,\pi)+N^{i}\mathcal{H}_{i}(\gamma,\pi)+\sqrt{\gamma}\lambda_{C}\!\left(\Mpl^{2}\,\phi-\frac{\pi^{ij}}{\sqrt{\gamma}}\,\gamma_{ij}\right) \nonumber \\
 &  & {}+\lambda_{\phi}\pi_{\phi}+\sqrt{\gamma}\Mpl^{2}\lambda_{{\rm gf}}^{i}\,\partial_{i}\phi+N\sqrt{\gamma}\Mpl^{2}V(\phi)\bigg]\,, \label{eqn:Htot}
\end{eqnarray}
where we consider $N,N^{i}$ as Lagrange multipliers. Obviously, we recover General Relativity in the $V\to{\rm const}$ limit.

\section{Lagrangian formulation}
\label{sec:Lagrangian}

The Lagrangian can be found by performing a Legendre transformation. For this purpose we calculate $\dot{\gamma}_{ij}$ as follows
\begin{equation}
\dot{\gamma}_{ij}=\{\gamma_{ij},H_{{\rm tot}}\}=\frac{2N}{\Mpl^{2}}\,\frac{\pi^{lm}}{\sqrt{\gamma}}(2\gamma_{il}\gamma_{jm}-\gamma_{lm}\gamma_{ij})+D_{i}N_{j}+D_{j}N_{i}-\lambda_{C}\,\gamma_{ij}\,.
\end{equation}
If we use the definition for the extrinsic curvature, namely
\begin{equation}
K_{ij}\equiv\frac{1}{2N}\,(\dot{\gamma}_{ij}-D_{i}N_{j}-D_{j}N_{i})\,,
\end{equation}
then we obtain 
\begin{equation}
2NK_{ij}=\frac{2N}{\Mpl^{2}}\,\frac{\pi^{lm}}{\sqrt{\gamma}}(2\gamma_{il}\gamma_{jm}-\gamma_{lm}\gamma_{ij})-\lambda_{C}\,\gamma_{ij}\,,
\end{equation}
so that
\begin{eqnarray}
2NK_{ij}(\gamma^{ir}\gamma^{js}-\gamma^{rs}\gamma^{ij}) & = & \frac{2N}{\Mpl^{2}}\,\frac{\pi^{lm}}{\sqrt{\gamma}}(2\gamma_{il}\gamma_{jm}-\gamma_{lm}\gamma_{ij})(\gamma^{ir}\gamma^{js}-\gamma^{rs}\gamma^{ij}) \nonumber \\
 &  & {}-\lambda_{C}\,\gamma_{ij}(\gamma^{ir}\gamma^{js}-\gamma^{rs}\gamma^{ij})\nonumber \\
 & = & \frac{4N}{\Mpl^{2}}\,\frac{\pi^{rs}}{\sqrt{\gamma}}+2\lambda_{C}\gamma^{rs}\,,
\end{eqnarray}
or
\begin{equation}
\frac{\pi^{ij}}{\sqrt{\gamma}}=\frac{\Mpl^{2}}{2}\left(K^{ij}-K\gamma^{ij}-\frac{\lambda_{C}\gamma^{ij}}{N}\right)\,.
\end{equation}

After a suitable redefinition of the Lagrange multiplier $\lambda_{C}=N\lambda$, we obtain the following Lagrangian
\begin{eqnarray}
  \mathcal{L}=N\sqrt{\gamma} &  & \Bigg[\frac{\Mpl^{2}}{2}\left(R+K_{ij}\,K^{ij}-K^{2}-2V(\phi)\right) \nonumber \\
  &  & \quad - \frac{\lambda_{{\rm gf}}^{i}}{N}\,\Mpl^{2}\,\partial_{i}\phi-\frac{3\Mpl^{2}\lambda^{2}}{4}-\Mpl^{2}\lambda\,(K+\phi)\Bigg]\,.
  \label{eq:Lagr}
\end{eqnarray}
Here we consider $[\phi]=M,[\lambda]=M$, $[V]=M^{2}$.  As already stated above, General Relativity is recovered in the limit $V_{,\phi}\to0$. To this Lagrangian we can now add the standard matter Lagrangian terms minimally coupled to the metric~(\ref{eqn:metric-ADM}).

\section{Number of degrees of freedom}
\label{sec:degrees}
Let us consider $N$, and $N^{i}$ as Lagrange multipliers in the defining Hamiltonian for the theory (\ref{eqn:Htot}). In the following, we find it useful to introduce for any scalar density $\mathcal{C}$ and vector density $\mathcal{C}^i$ (or $\mathcal{C}_i$) a test scalar function $v$ and a test vector $v_i$ (or $v^i$) so that we can build the following integrals 
\begin{equation}
\left\langle \mathcal{C}\right\rangle _{v}\equiv\int d^{3}x\,v\,\mathcal{C}\,, \quad
\left\langle \mathcal{C}^i\right\rangle _{v_i}\equiv\int d^{3}x\,v_i\,\mathcal{C}^i\,, \quad
\left(\mbox{ or } \,  \left\langle \mathcal{C}_i\right\rangle _{v^i}\equiv\int d^{3}x\,v^i\,\mathcal{C}_i\,  \right)\,. 
\end{equation}
Let us also define the following constraints
\begin{eqnarray}
C_{1} & = & \sqrt{\gamma}\left[\Mpl^{2}V-\frac{\Mpl^{2}}{2}\,R+\frac{1}{\Mpl^{2}}\,(2\Pi_{ij}\,\Pi^{ij}-\tilde{\pi}^{2})\right]\,,\\
C_{2} & = & \sqrt{\gamma}\Pi_{\phi}\,,\\
C_{3} & = & \sqrt{\gamma}\left(\Mpl^{2}\phi-\frac{\pi^{ij}}{\sqrt{\gamma}}\,\gamma_{ij}\right)\,,\\
C_{4}^{i} & = & -2\sqrt{\gamma}\,\nabla_{j}\Pi^{ij}\,,\\
C_{5i} & = & \sqrt{\gamma}\,\Mpl^{2}\,\partial_{i}\phi\,,
\end{eqnarray}
where a tilde means a division by $\sqrt{\gamma}$. It follows from $C_{3}\approx0$ that 
\begin{equation}
\tilde{\pi}\equiv\frac{\pi^{ij}}{\sqrt{\gamma}}\,\gamma_{ij}\,,
\end{equation}
is $\Mpl^2\phi$ on the constraint surface. Therefore, on the constraint surface, we also have
\begin{equation}
\nabla_{i}\tilde{\pi}\approx \Mpl^2\partial_{i}\phi\approx0\,,
\end{equation}
where we have used $C_{5i}\approx0$. We can build a set of first class constraints, ``extended $C_{4}^{i}$'' as follows
\begin{equation}
C_{4E}^{i}=-2\sqrt{\gamma}\,\nabla_{j}\Pi^{ij}+\sqrt{\gamma}\,\Pi_{\phi}\,\nabla^{i}\phi\,.
\end{equation}
Indeed these expressions have weakly vanishing Poisson brackets with any other constraints as well as with themselves, as shown in appendix~\ref{sec:appo}.

Let us look at the secondary constraints. Following the calculations given in appendix~\ref{sec:appo},  we find that on setting
\begin{equation}
\{\left\langle C_{1}\right\rangle _{v_{1}},H_{{\rm tot}}\}\approx0\,,
\end{equation}
we find an expression which can be solved for $\lambda_{C}$. As for the constraint
\begin{equation}
\{\left\langle C_{2}\right\rangle _{v_{2}},H_{{\rm tot}}\}\approx0\,,
\end{equation}
it fixes the quantity $\nabla_{i}\lambda_{{\rm gf}}^{i}$. Also, one can show that
\begin{equation}
\{\left\langle C_{3}\right\rangle _{v_{3}},H_{{\rm tot}}\}\approx0\,,
\end{equation}
can be solved for $N$. On the other hand, one can verify that the relation
\begin{equation}
\{\left\langle C_{4E}^{i}\right\rangle _{v_{4i}},H_{{\rm tot}}\}\approx0\,,
\end{equation}
is satisfied automatically, so that it does not add any new constraint.
Finally on imposing
\begin{equation}
\{\left\langle C_{5}^{i}\right\rangle _{v_{5i}},H_{{\rm tot}}\}\approx0\,,
\end{equation}
we find that $\nabla_{i}\lambda_{\phi}\approx0$, which fixes $\lambda_{\phi}$,
and does not add any new constraint. So we do not find any new secondary
constraint to impose. For more details see appendix~\ref{sec:appo}.

We can now count the number of degrees of freedom. In total we have $2\times6+2=14$ phase space variables (6 from $\gamma_{ij}$, 6 from $\pi^{ij}$, 1 from $\phi$ and 1 from $\pi_{\phi}$). We also have 3 first-class and 4 second-class constraints, so that we end up with $14-2\times3-1\times4=4$ phase space variables, or 
2 physical degrees of freedom. We have counted the constraints $C_{5}^{i}$
as only one, as indeed by integration by parts $\left\langle C_{5i}\right\rangle_{\lambda_{\rm gf}^i}$ can be shown to be proportional to $\left\langle \nabla^{2}\phi\right\rangle_{\lambda_{{\rm gf},S}}$, where we have decomposed without loss of generality $\lambda_{\rm gf}^i=\gamma^{ij}\,\partial_j\lambda_{{\rm gf},S}+\lambda_{{\rm gf},V}^i$, with $D_i\lambda_{{\rm gf},V}^i=0$.

\section{Cosmology}
\label{sec:cosmology}
In the following we will study the behavior of the theory introduced above on a flat Friedmann-Lema\^\i tre-Robertson-Walker (FLRW) background. This will help us also considering the nature of the propagating modes, and the influence on the evolution for both the background and perturbation variables.

In order to study the flat FLRW background and the perturbations, we will make use of the Lagrangian of the theory defined in eq.\ (\ref{eq:Lagr}). For an analogue study of the background via a mini-super-space Hamiltonian, one can refer to the results given in appendix \ref{sec:mini_ham}. We will add some matter field minimally coupled to the metric (\ref{eqn:metric-ADM}), in the form of a perfect fluid with the barotropic equation of state $P=P(\rho)$ (see e.g. \cite{Pookkillath:2019nkn} and references therein for details), to the basic Lagrangian of eq.\ (\ref{eq:Lagr}).

\subsection{Background equations of motion}

We consider the flat FLRW background metric 
\begin{equation}
 N = N(t)\,, \quad N^i = 0\,, \quad \gamma_{ij} = a^2(t)\delta_{ij}\,,
\end{equation}
and the homogeneous matter energy density $\rho=\rho(t)$. We find that the equation of motion for the field $\lambda$ gives
\begin{equation}
  \lambda=-\frac23\,\phi-2H\,,
\end{equation}
where $H$ is the Hubble parameter, i.e.\ $H\equiv \dot{a}/(aN)$. 

The equation of motion for the lapse, i.e.\ the first Einstein equation, gives
\begin{equation} 
  \phi^2=3V+\frac{3\rho}{\Mpl^2}\,.\label{eq:Einstein1}
\end{equation}
On taking a linear combination of the second and the first Einstein equations, we find
\begin{equation}
  \frac{\dot\phi}{N}=\frac32\,\frac{\rho+P}{\Mpl^2}\,.\label{eq:dot_phi}
\end{equation}
The equation of motion for the field $\phi$ gives
\begin{equation}
  \phi=\frac32\,V_{,\phi}-3H\,,\label{eq:phieq}
\end{equation}
whereas the matter fields obey the standard conservation equation
\begin{equation}
  \frac{\dot\rho}N+3H(\rho+P)=0\,.
\end{equation}

On combining the first Einstein equation (\ref{eq:Einstein1}) and the $\phi$-equation of motion (\ref{eq:phieq}), it is possible to rewrite the Friedmann equation in a more familiar form as follows
\begin{equation}
3\Mpl^{2}H^{2}=\rho+\rho_{\phi}\,, \label{eqn:Friedmann-eq}
\end{equation}
where
\begin{equation}
\rho_{\phi}\equiv\Mpl^{2}(V-\phi V_{,\phi})+\frac{3}{4}\,\Mpl^{2}\,V_{,\phi}^{2}\,.
\end{equation}
On taking derivatives of the eqs.\ (\ref{eq:dot_phi}) and (\ref{eq:phieq}), we can find expressions for the quantities $\ddot\phi$ and $\ddot{a}$ respectively. In particular, one finds that
\begin{equation}
\frac{\dot{H}}{N}=\frac{ ( \rho+P )\,( 3\,V_{{,\phi\phi}}-2 ) }{4\Mpl^2}\,,\label{eq:dot_H}
\end{equation}
which deviates from the corresponding equation in GR if and only if $V_{,\phi\phi} \ne 0$. Under the same condition, (\ref{eqn:Friedmann-eq}) shows deviation from GR with a cosmological constant. On studying Eq.\ (\ref{eq:dot_H}), we find that we can express it in terms of an effective pressure defined by
\begin{equation}
  P_\phi=-\frac32\,(\rho+P)\,V_{,\phi\phi}-\rho_\phi\,,
\end{equation}
so that, we can find the effective equation of state for such a component as
\begin{eqnarray}
  w_\phi&=&\frac{P_\phi}{\rho_\phi}=-1-\frac32\,\frac{\rho+P}{\Mpl^2}\,\frac{V_{,\phi\phi}}{V-\phi V_{,\phi}+\frac{3}{4}\,V_{,\phi}^{2}}\nonumber\\
  &=&-1-\frac92\,(1+w)\,\Omega\,\frac{V_{,\phi\phi}(\frac12 V_{,\phi}-\frac\phi3)^2}{V-\phi V_{,\phi}+\frac{3}{4}\,V_{,\phi}^{2}}\,,
  \end{eqnarray}
where $w=P/\rho$, and $\Omega=\rho/(3\Mpl^2H^2)$.

\subsection{Reconstructing the potential}

Let us consider the problem of reconstructing the potential $V(\phi)$ for a given background dynamics. In terms of the e-fold variable,  $\mathcal{N=\ln}(a/a_{0})$, the set of independent background equations of motion is
\begin{eqnarray}
 V & = & \frac{1}{3}\phi^2 -\frac{\rho}{M_{\rm Pl}^2}\,, \label{eqn:V}\\
 \frac{d\phi}{d\mathcal{N}} & = & \frac{3}{2}\frac{\rho+P}{M_{\rm Pl}^2H}\,,\label{eqn:dphidN}\\
\frac{d\rho_i}{d\mathcal{N}} & = & -3(\rho_i + P_i)\,, 
\end{eqnarray}
where $\rho=\sum_i\rho_i$ and $P=\sum_iP_i$. Unless $\rho+P=0$, the following equation follows from the above equations:
\begin{equation} \label{eq:H_eq}
\phi = \frac{3}{2}V_{,\phi} - 3H\,. 
\end{equation}

From Eq.\ (\ref{eqn:dphidN}), one obtains
\begin{equation}
 \phi(\mathcal{N}) = \phi_0 + \int_{\mathcal{N}_0}^{\mathcal{N}}\frac{3}{2}\frac{\rho(\mathcal{N}')+P(\mathcal{N}')}{M_{\rm Pl}^2H(\mathcal{N}')}d\mathcal{N}'\,, \label{eqn:phi}
\end{equation}
where $\phi_0=\phi(\mathcal{N}_0)$. Assuming that 
\begin{equation}
 \rho + P > 0\,, \quad H > 0\,, \label{eqn:assumption-for-reconstruction}
\end{equation}
the right hand side of (\ref{eqn:phi}) is an increasing function of $\mathcal{N}$ and thus the function $\phi(\mathcal{N})$ has a unique inverse function,
\begin{equation}
 \mathcal{N} = \mathcal{N}(\phi)\,.
\end{equation}
Obviously, $\mathcal{N}$ is an increasing function of $\phi$. By combining this with (\ref{eqn:V}), one obtains
\begin{equation}
 V(\phi) = \frac{1}{3}\phi^2 - \frac{\rho(\mathcal{N}(\phi))}{M_{\rm Pl}^2}\,.
\end{equation}
The assumption (\ref{eqn:assumption-for-reconstruction}) simply states that at the level of the FLRW background, the total stress-energy tensor of matter fields should satisfy the null energy condition and that the universe should be expanding. Since the assumption does not involve the auxiliary field $\phi$ or the potential $V(\phi)$, the reconstruction of the potential works for any $\rho_{\phi}(\mathcal{N})$ and thus for any $w_{\phi}(\mathcal{N})$ as far as they do not stop the expansion of the universe. 

Just to show an example, let us consider the case that of discussing an accelerating dynamics, where we suppose this phenomenon happens at very late times, for which
\begin{equation}
q<0\,,\qquad{\rm where}\qquad-\frac{H'}{H}=1+q\,.
\end{equation}
On considering a $\mathcal{N}$-derivative of eq.\ (\ref{eq:H_eq}),
we have, at late times
\begin{equation}
q=-1-\frac{H'}{H}=-1-\frac{(3V_{,\phi\phi}-2)(\rho+P)}{4\Mpl^{2}H^{2}}\approx-1-\frac{3}{4}\,(3V_{,\phi\phi}-2)\,\Omega_{m}\,.
\end{equation}
We have acceleration provided that 
\begin{equation}
V_{,\phi\phi}|_{0} \gtrsim -\frac{22}{27}\,,
\end{equation}
where we have assumed $\Omega_{m0}\approx0.3$ and $w\approx0$.

\subsection{Tensor perturbation}

We have studied the background dynamics, and now we want to study the propagation of the tensor modes. In fact, on considering the tensor degrees of freedom, namely
\begin{equation}
ds^{2}=-N^{2}\,dt^{2}+a^{2}\,\delta_{ij}\,dx^{i}\,dx^{j}+a^{2}\sum_{\sigma=+,\times}h_{\sigma}\,\epsilon_{\mu\nu}^{(\sigma)}\,dx^{\mu}\,dx^{\nu}\,,
\end{equation}
for which the two polarization tensors, $\epsilon_{\mu\nu}^{(\sigma)}$,
are normalized to unity, then we find that the action for the tensor
modes reduces to
\begin{equation}
S=\frac{\Mpl^{2}}{8}\int dtd^{3}x\,Na^{3}\sum_{\sigma}\left[\frac{\dot{h}_{\sigma}^{2}}{N^{2}}-\frac{1}{a^{2}}\,\delta^{ij}(\partial_{i}h_{\sigma})(\partial_{j}h_{\sigma})\right]\,.
\end{equation}
This is independent of the field $\phi$ and agrees with the corresponding expression in GR. Although we do not show it explicitly, the vector modes also reduce to exactly the same dynamics of GR.

\subsection{Scalar perturbation}

If we look for the stability and degrees of freedom on a flat FLRW universe for the scalar sector we need to perturb the metric elements as follows
\begin{eqnarray}
  N &=& N(t)(1+\alpha)\,,\\
  N_i&=&N(t)\partial_i\chi\,,\\
  \gamma_{ij}&=&a^2\,(1+2\Phi)\,\delta_{ij}\,,
\end{eqnarray}
where we have fixed the 3D-diffeomorphism invariance by setting a gauge for which the 3D metric $\gamma_{ij}$ is diagonal. Furthermore, for this theory, we also need the following expansion of fields 
\begin{eqnarray}
  \phi&=&\phi(t)+\delta\phi\,,\\
  \lambda&=&\lambda(t)+\delta\lambda\,,\\
  \lambda^i_{\rm gf}&=&\frac1{a^2}\,\,\delta^{ij}\,\partial_j\delta\lambda_{\rm gf}\,,
\end{eqnarray}
whereas the standard matter field is treated as a perfect fluid following the action and variables introduced in \cite{Pookkillath:2019nkn}. In particular we will call by
$v_m$ the scalar part of the 3D component of the fluid 4-velocity, that is $u_i=\partial_iv_m$, and by $\delta\rho$ the perturbation of the fluid energy density.

By expanding the Lagrangian density at second order in the perturbation variables, on integrating out all the auxiliary fields, we find that only one scalar mode is propagating (one for each of the standard matter species), as expected. It is possible to write the so obtained reduced Lagrangian density in terms of a gauge invariant variable, namely
\begin{equation}
  \delta_m\equiv\frac{\delta\rho}{\rho}-\frac{3H(\rho+P)}{\rho}\,v_m\,,
\end{equation}
and we find that in the high-$k$ regime, the action reduces
to the standard results of GR, namely the squared speed of propagation becomes
\begin{equation}
c_{s}^{2}=\left(\frac{\partial P}{\partial\rho}\right)_{s}\,,
\end{equation}
and the no-ghost condition (for high $k$) assumes the standard form, namely: $Q_{s}>0$, where
\begin{equation}
Q_{s}=\frac{a^{2}\,\rho^{2}}{2k^{2}(\rho+P)}\,.
\end{equation}

Furthermore, it is possible to rewrite the reduced Lagrangian for the perturbation variable $\delta_m$ so that its coefficients (the ones of $\delta_m^{2}$ and $\dot{\delta}_m^{2}$) do not explicitly depend on $\phi$ and its derivatives, but only on $\dot{H}$, $H$ and the matter variables (in addition of course to the modulus of the wave vector, $k$). This fact is shown explicilty in appendix \ref{sec:app_baro}. However, on replacing $\dot{H}$ by the expression given in eq.\ (\ref{eq:dot_H}), the dependence on $V_{,\phi\phi}$ will reappear. Therefore the dynamics of $\delta_m$ will be in general different from the one of GR.

As an example, we will from now on focus on a pressure-less fluid, i.e.\  $P=P_m=0$ and $\rho=\rho_m\propto a^{-3}$, finding that $c_{s}^{2}=0$, as expected. In terms of the gauge invariant variable $\delta_m$, the reduced action for the scalar perturbation can be written as
\begin{equation}
  S = \frac12\int d^4x\, Na^3\left[ \frac{ \left( \frac32\,\rho_m+\frac{{k}^{2}}{a^2}\Mpl^2 \right)a^2 \rho_m\,{\dot\delta}_m^2}
        {\left(\frac32\rho_m+\frac{{k}^{2}}{a^2}\Mpl^2  -\frac94\,\rho_mV_{{,\phi\phi}}\right) {k}^{2}N^2}+\frac{a^2\rho_m^2\delta_m^2}{2k^2\Mpl^2}\right].
\end{equation}
This expression, due to the presence of $V_{,\phi\phi}$, leads to corrections to the dynamics of clustering of the matter perturbation, i.e.\ a change in the effective gravitational constant $G_{{\rm eff}}/G_{N}$. However, at large $k$'s (or whenever $|V_{,\phi\phi}|\ll1$ holds), GR is recovered, i.e. $G_{\rm eff}/G_{N} \to 1$. In particular deviations from GR take place both at the mass and the friction terms as it can be seen in the following equation of motion 
\begin{eqnarray}
  \delta_m''&+&\frac1{4\Delta}\,\{486H{\Omega_{m}^2} (2 {K}^{2}+9\,\Omega_{{m}}) V_{{,\phi\phi\phi}}\nonumber\\
            &+&36\,\Omega_{{m}} ( 2\,{K}^{4}+27\,{K}^{2}\Omega_{{m}}-18\,{K}^{2}+81\,{\Omega_{m}^2}-54\,\Omega_{{m}} ) V_{{,\phi\phi}}-243\Omega_m^2( 2\,{K}^{2}+9\,{\Omega_{m}} ) {V_{{,\phi\phi}}^2}\nonumber\\
  &-&4( 2\,{K}^{2}+9\,\Omega_{m} ) ^{2} ( 3\,\Omega_{{m}}-4 )\}\,\delta_m'
+{\frac {27\,{\Omega_{m}^2} ( 3\,V_{{,\phi\phi}}-2 )-12\,{K}^{2}\Omega_{{m}} }{8\,{K}^{2}+36\,\Omega_{{m}}}}\,\delta_m=0\,,\\
  \Delta&\equiv&( 2\,{K}^{2}+9\,\Omega_{{m}})(4\,{K}^{2}-27\,\Omega_{{m}}V_{{,\phi\phi}}+18\,\Omega_{{m}}) \,,
\end{eqnarray}
where we have introduced $K\equiv k/(aH)$, $\Omega_m=\rho_m/(3\Mpl^2 H^2)$, and a prime denotes differentiation with respect to the e-fold variable $\mathcal{N}=\ln(a/a_0)$. Notice that, because of the non-trivial background modification, the term $V_{,\phi\phi}$ appears in the high $k$ regime in the friction term. Therefore a non-trivial phenomenology arises for this kind of theory, which can in principle attempt to solve today's puzzles in cosmology.

\section{Summary and discussion}
\label{sec:summary}
We have introduced a type-II minimally modified gravity (MMG) theory, that is a theory with only two degrees of freedom (in vacuum), as in General Relativity (GR), which does not possess an Einstein frame. Our initial aim was to be able to implement a dark energy component in the energy budget of the universe which, however, does not introduce any new degree of freedom, and which could lead to not large modifications to standard matter fields, e.g.\ radiation or baryon fields. This picture, if implemented, makes it possible, in our aim, to possibly avoid problems related to the stability of the background which typically arise when we want to achieve some non-trivial background/perturbation behavior. In fact, we do not need to worry about the possibility of the new matter mode to become a ghost, simply because there is no extra gravitational mode besides the standard tensor modes.

To reach this goal, we have started building this theory from the GR Hamiltonian and have performed a canonical transformation (which is invertible) to another arbitrary frame. So far the new theory is still intrinsically GR, although written by means of other variables. However, in the new frame we add two new bits: 1) a cosmological constant, and 2) a gauge-fixing term. The first term breaks the equivalence between the theory under consideration and GR, whereas the second term is introduced to keep the number of degrees of freedom to be only two.

After having introduced this new-frame not-GR Hamiltonian, we go back to the original frame by means of the inverse canonical transformation but, this time, not to GR. We have then reached our goal: to introduce a dark energy component, whose dynamics in general depends on time, which, nonetheless, does not add any new degree of freedom.

We have then studied this theory, in the Hamiltonian formalism, in order to confirm that the number of degrees of freedom is two and only two on any background and at non-linear level. The effective cosmological constant in the other frame now becomes actually time dependent, leading to a non-trivial dark sector. Its size determines the scale at which these modification will affect the physics of the background and of the perturbation variables. Setting its size to values comparable to $H_0^2$ implies that such a modification can be safely neglected at high energies and at small scales. This is possible also because (standard) matter fields are still minimally coupled to gravity, i.e.\ they are not coupled directly to this source of the cosmic acceleration. 

We have then extended the study of this theory to the propagation of the perturbation variables about a general Friedmann-Lema\^\i tre-Robertson-Walker (FLRW) background. Indeed we find only the tensor modes to be propagating in vacuum. On top of that we find that, on introducing standard matter degrees of freedom, this theory can affect the evolution of the matter perturbation variables, but only at late times and at large scales, potentially leading to an interesting and ghost-free phenomenology.

At the beginning of Sec~\ref{sec:intro} we have listed six steps for the construction of a type-II MMG theory. If in step 3 we added matter fields together with the cosmological constant in the same new frame, then the theory would have been the type-I MMG theory studied in \cite{Aoki:2018brq}. Also, if we did not add a cosmological constant in the new frame, i.e. if we skipped step 3, then we would have ended up with gauge-fixed GR after adding matter fields to the original frame in step 6. On the other hand, for the construction of the type-II MMG theory, we add a cosmological constant in the new frame and matter fields in the original frame. This way we obtain a MMG theory which is different from the one studied in \cite{Aoki:2018brq}. This also explains the reason why this theory should be of type-II: neither the new frame after step 2 nor the original frame after step 5 is an Einstein frame.

Finally, we want to mention that this same mechanism can be extended to other dark sectors. In particular, this same model-building algorithm can be applied to a dark matter component or to several other combined dark sectors. In principle different dark sector components can be coupled to different frames that are generated from the common original frame by different canonical transformations. In the future, we want to use this theory to be able to implement non-trivial behaviors, such as weak gravity and to address the problems which affect at the moment the standard model of cosmology.

\appendix
\section{Useful relations}
\label{sec:appo}
In order to prove the statements made in Section~\ref{sec:degrees}, it turns out that the following
relations, valid for any scalar $A$, vector $V^{i}$ and tensor $M^{ij},$
can be useful
\begin{eqnarray}
\nabla_{i}\nabla_{j}\nabla^{j}A & = & \nabla_{j}\nabla^{j}\nabla_{i}A-R_{ij}\nabla^{j}A\,,\\
V^{j;i}{}_{;j} & = & V^{j}{}_{;j}{}^{;i}+R^{i}{}_{j}V^{j}\,,\\
V^{j;k}{}_{;i} & = & V^{j}{}_{;i}{}^{;k}+R^{j}{}_{lim}\,g^{km}\,V^{l}\,,\\
V^{i}{}_{;i}{}^{;j}{}_{;j}-V^{i;j}{}_{;j;i} & = & V^{i}{}_{;jkl}\,(\gamma^{kl}\delta^{j}{}_{i}-\gamma^{jk}\delta^{l}{}_{i})\nonumber \\
 & = & -\frac{1}{2}\,V^{i}\nabla_{i}R-\nabla^{i}V^{j}\,R_{ij}\,,\\
M_{i}{}^{k}{}_{;jk} & = & M_{i}{}^{k}{}_{;kj}+R_{kj}\,M_{i}{}^{k}+R^{l}{}_{ijk}\,M_{l}{}^{k}\,,
\end{eqnarray}
where we have used the relation $\Gamma^{l}{}_{ki,j}-\Gamma^{l}{}_{kj,i}=R^{l}{}_{kji}$,
valid in a local inertial frame.

Then we find
\begin{equation}
\{\left\langle C_{5i}\right\rangle _{v_{5}^{i}},H_{{\rm tot}}\}_{{\rm PB}}\approx\int d^{3}x\sqrt{\gamma}\,v_{5}^{i}\,\Mpl^{2}\,\partial_{i}\lambda_{\phi}\approx0\,,
\end{equation}
which can be solved for $\lambda_{\phi}$. Then we also have
\begin{eqnarray}
\{\left\langle C_{1}\right\rangle _{v_{1}},H_{{\rm tot}}\}_{{\rm PB}} & \approx & -\int d^{3}x\sqrt{\gamma}v_{1}\left\{ \frac{\lambda_{C}(\Mpl^{4}V+\tilde{\pi}^{2}-4\tilde{\pi}_{ij}\tilde{\pi}^{ij})}{\Mpl^{2}}\right.\nonumber \\
 & + & \frac{N^{i}}{2\Mpl^{2}}\left(\Mpl^{4}\nabla_{i}R-8\tilde{\pi}^{jk}\nabla_{i}\tilde{\pi}_{jk}\right)+\Mpl^{2}\,\nabla_{i}\nabla^{i}\lambda_{C}\nonumber \\
 & - & \left.\Mpl^{2}\tilde{\Lambda}\lambda_{\phi}\,V_{,\phi}\right\} ,
\end{eqnarray}
which can be solved in principle for $\lambda_{C}$, since $\lambda_{\phi}$
has been already fixed. In the following, a prime denotes derivative with respect
to $\phi$. For the constraint $C_{2}$, we find
\begin{eqnarray}
\{\left\langle C_{2}\right\rangle _{v_{2}},H_{{\rm tot}}\}_{{\rm PB}}\approx\int d^{3}x\sqrt{\gamma}v_{2} & & \Bigg[\nabla_{i}\lambda_{{\rm gf}}^{i}+\frac{1}{2}\,\lambda_{C}\left(\frac{\tilde{\pi}f'_{0}}{\Mpl^{2}\sqrt{f_{0}}}-2\right) \nonumber \\
 & & \qquad\quad\, +\Mpl^{2}\tilde{\Lambda}\,N\,\frac{3f_{1}f'_{0}+2f_{0}f'_{1}}{2f_{0}^{5/2}f_{1}^{2}}\Bigg],
\end{eqnarray}
which now fixes $\nabla_{i}\lambda_{{\rm gf}}^{i}$. Once more this
is only one condition on the scalar part of $\lambda_{{\rm gf}}^{i}$.

Let us now calculate
\begin{equation}
\{\left\langle C_{3}\right\rangle _{v_{3}},H_{{\rm tot}}\}_{{\rm PB}}\approx\int d^{3}x\sqrt{\gamma}v_{3}\left[\Mpl^{2}\nabla_{i}\nabla^{i}N+N\,\frac{\Mpl^{4}V+\tilde{\pi}^{2}-4\tilde{\pi}_{ij}\tilde{\pi}^{ij}}{\Mpl^{2}}+\Mpl^{2}\,\lambda_{\phi}\right],
\end{equation}
which can be calculated in terms of $N$. Since $\{\left\langle C_{4E}^{i}\right\rangle _{v_{4i}},H_{{\rm tot}}\}\approx0$
is automatically satisfied, this shows that there are no secondary
constraints to add to the Hamiltonian.

Also we find the following constraint algebra
\begin{eqnarray}
\{\left\langle C_{1}\right\rangle _{v_{1}},\left\langle C_{2}\right\rangle _{v_{2}}\}_{{\rm PB}} & \approx & \int d^{3}x\sqrt{\gamma}v_{1}v_{2}\,\Mpl^{2}\,V_{,\phi}\,,\\
\{\left\langle C_{1}\right\rangle _{v_{1}},\left\langle C_{3}\right\rangle _{v_{3}}\}_{{\rm PB}} & \approx & \int d^{3}x\sqrt{\gamma}\sqrt{f_{0}}v_{3}\bigg[v_{1}\left(\frac{4\tilde{\pi}_{ij}\tilde{\pi}^{ij}-\tilde{\pi}^{2}}{\Mpl^{4}}-\Mpl^{2}V\right) \nonumber \\
 & & \qquad\qquad\qquad\qquad\qquad -\Mpl^{2}\nabla_{i}\nabla^{i}v_{1}\bigg],\\
\{\left\langle C_{1}\right\rangle _{v_{1}},\left\langle C_{4Ei}\right\rangle _{v_{4}^{i}}\}_{{\rm PB}} & \approx & 0\,,\\
\{\left\langle C_{1}\right\rangle _{v_{1}},\left\langle C_{5i}\right\rangle _{v_{5}^{i}}\}_{{\rm PB}} & \approx & 0\,,\\
\{\left\langle C_{2}\right\rangle _{v_{2}},\left\langle C_{3}\right\rangle _{v_{3}}\}_{{\rm PB}} & \approx & -\Mpl^{2}\int d^{3}x\sqrt{\gamma}\,v_{2}v_{3},\\
\{\left\langle C_{2}\right\rangle _{v_{2}},\left\langle C_{4Ei}\right\rangle _{v_{4}^{i}}\}_{{\rm PB}} & \approx & 0\,,\\
\{\left\langle C_{2}\right\rangle _{v_{2}},\left\langle C_{5i}\right\rangle _{v_{5}^{i}}\}_{{\rm PB}} & \approx & \Mpl^{2}\int d^{3}x\sqrt{\gamma}\,v_{2}\nabla_{i}v_{5}^{i}\,,\\
\{\left\langle C_{3}\right\rangle _{v_{3}},\left\langle C_{4Ei}\right\rangle _{v_{4}^{i}}\}_{{\rm PB}} & \approx & 0\,,\\
\{\left\langle C_{3}\right\rangle _{v_{3}},\left\langle C_{5i}\right\rangle _{v_{5}^{i}}\}_{{\rm PB}} & \approx & 0\,,\\
\{\left\langle C_{4Ei}\right\rangle _{v_{4}^{i}},\left\langle C_{5i}\right\rangle _{v_{5}^{i}}\}_{{\rm PB}} & \approx & 0\,.
\end{eqnarray}
This algebra shows that $C_{4Ei}$ represent three first class constraints. Moreover,
we do not find any other first class combination of constraints, or
any other necessary secondary constraints.

\section{Mini-super-space Hamiltonian}
\label{sec:mini_ham}

It can be proven that in minisuperspace the Hamiltonian density can be written
as
\begin{equation}
H=\bar{\lambda}_{\phi}\pi_{\phi}-\frac{\pi_{a}^{2}N}{12\Mpl^{2}a}+a^{3}\bar{\lambda}_{C}\left(\Mpl^{2}\phi-\frac{\pi_{a}}{2a^{2}}\right)+Na^{3}V\Mpl^{2}\,,
\end{equation}
for which we have three primary constraints determined by the three
different Lagrange multipliers, and we call them
\begin{eqnarray}
C_{1} & = & \Mpl^{2}a^{3}V-\frac{\pi_{a}^{2}}{12\Mpl^{2}a}\,,\\
C_{2} & = & \pi_{\phi}\,,\\
C_{3} & = & a^{3}\left(\Mpl^{2}\phi-\frac{\pi_{a}}{2a^{2}}\right).
\end{eqnarray}
Then we have
\begin{equation}
\{C_{2},H\}=-\Mpl^{2}a^{3}[NV_{,\phi}+\bar{\lambda}_{C}]\approx0\,,
\end{equation}
which in general fixes $\bar{\lambda}_{C}$, so that we do not obtain
a new constraint. Notice that in the GR limit, $V_{,\phi}\to0$, in
general we find that $\bar{\lambda}_{C}\to0$. Finally we find
\begin{equation}
\{C_{3},H\}={\it M_{\rm P}}^{2}a^{3}\bar{\lambda}_{\phi}+\frac{N\left(12\,\Mpl^{4}a^{4}V-4\,\Mpl^{2}a^{2}\phi\,\pi_{{a}}+\pi_{{a}}^{2}\right)}{8\Mpl^{2}a}\approx0\,,
\end{equation}
which actually sets the Lagrangian multiplier $\bar{\lambda}_{\phi}$.
On substituting these relations in $\{C_{1},H\}$, then we find it
vanishes identically. Therefore we do not obtain any new constraint.
Therefore the total Lagrangian becomes $H_{{\rm tot}}=H$.

Since $\{C_{1},C_{2}\}\neq0$, $\{C_{1},C_{3}\}\neq0$, and $\{C_{2},C_{3}\}\neq0$,
then $C_{1},C_{2},C_{3}$ are second class constraints.

In this mini-super-space we have 4 phase-space variables, namely $a,\phi$
and their conjugate momenta. And we have $4-1\times3=1$ degree of
freedom left for the background, for which the gauge-fixing constraint
does not give contributions.

\section{Perturbation of a general barotropic perfect fluid}
\label{sec:app_baro}
In this appendix we write down explicitly the quadratic action of the scalar perturbation for a general barotropic fluid with equation of state $\rho=\rho(n)$, and $P=P(n)=n\rho_{,n}-\rho$, so that $P=P(\rho)$, where $\rho$ is the fluid energy density, $n$ its number density (which is proportional to $a^{-3}$) and $P$ its pressure.

The reduced action will be written in terms of the gauge invariant variable
\begin{equation}
  \delta_m\equiv\frac{\delta\rho}{\rho} - \frac{3Hn\rho_{,n}}{\rho}\,v_m\,,
\end{equation}
where $n\rho_{,n}=\rho+P$, from the first principle of thermodynamics.
The important thing to notice here is the following. For the theory at hand, on using the equations of motion, it is possible to hide all the explicit dependence of $\phi$, its derivatives, and of $V$ and its derivatives in terms of $H$, its derivatives, and $\rho$ and its $n$-derivatives. What is interesting is that in this form, if we replace the standard GR background equations of motion the action reduces exactly to the one obtained in GR. Therefore, the two actions, once the one of the MMG is written in this form, are equivalent on GR-shell (i.e.\ when we rewrite $H$, $\dot{H}$, $\rho$ or its $n$-derivatives by using the GR background equations of motion).

In fact, this is how we will write such a reduced action. This reduced action is obtained after integrating out all the auxiliary fields. Such an action can be written schematically as
\begin{equation}
  S=\int d^4x[Q\,{\dot\delta}_m^2-W\,\delta_m^2]\,,
\end{equation}
where the coefficients $Q$ and $W$ can be written as 
\begin{eqnarray}
  Q&=&{\frac {3\Omega\,{a}^{3}\Mpl^2 \left( 2\,{K}^{2}+9\,w
\Omega+9\,\Omega \right) }{4N \left( {K}^{2}+3\,\epsilon \right) {K}^{2
} \left( 1+w \right) }}\,,\\
  W&=&{\frac {27\,N{H}^{2}\Omega\,\Mpl^2{a}^{3}}{2\, \left( {K}^{2}
       +3\,\epsilon \right) ^{2} \left( 1+w \right) {K}^{2}}}\left\{
{K}^{6}{c_{{s}}}^{2}+ \left[  \left( 6\,{c_{{s}}}^{2}+3\,w \right) 
\epsilon-\frac32 ( 1+w ) \Omega+9\,{c_{{s}}}^{2}-15\,w
       \right] {K}^{4}\right.\nonumber\\
  &+&\Bigg(  9( {c_{{s}}}^{2}-w ) {
\epsilon}^{2}+ \left[  \left( {\frac {27\,{w}^{2}}{2}}+\frac92\,w-9
 \right) \Omega+ 9( \eta-3 ) w+27\,{c_{{s}}}^{2}
 \right] \epsilon \nonumber \\
 &  & \quad +{\frac { 81( 1+w )  \left( w{c_{{s}}}^
     {2}-\frac43\,w+{c_{{s}}}^{2} \right) \Omega}{2}} \Bigg) {K}^{2}\nonumber\\
  &+&\left.{\frac {
243\,\epsilon\, \left( 1+w \right) \Omega\, \left(  \left( -w/3-1/9
 \right) \epsilon+ \left( {c_{{s}}}^{2}+\eta/3-2/3 \right) w+{c_{{s}}}
^{2} \right) }{2}}\right\}\,.
\end{eqnarray}
Here, 
\begin{equation}
  w = \frac{P}{\rho}\,,\quad 
  c_s^2 = \frac{n\rho_{,nn}}{\rho_{,n}}\,\quad 
  \epsilon = -\frac{\dot H}{NH^2}\,,\quad
  \eta = \frac{\dot\epsilon}{\epsilon N H}\,,\quad
  K = \frac{k}{aH}\,,\quad
  \Omega = \frac{\rho}{3\Mpl^2H^2}\,,
\end{equation}
and $w$ does not need to be constant. Once more, if we replace into this action the GR equations of motion, e.g.\ $\dot{H}=-N\rho(1+w)/(2\Mpl^2)$, we would obtain the expression for the quadratic action for the scalar perturbation in GR. However, in general the two theories are different, because of the different background dynamics for $\dot H$.

\acknowledgments
A.D.\ thanks Yukawa Institute for Theoretical Physics for their hospitality and provision of everything necessary for the research.
The work of A.D.F.\ \ was supported by Japan Society for the Promotion of Science Grants-in-Aid for Scientific Research No.~20K03969. The work of S.M.\ was supported by Japan Society for the Promotion of Science Grants-in-Aid for Scientific Research No.~17H02890, No.~17H06359, and by World Premier International Research Center Initiative, MEXT, Japan.

\end{document}